\documentclass[aps,prl,twocolumn,superscriptaddress]{revtex4-2}
\UseRawInputEncoding
\usepackage{graphicx}
\usepackage[caption=false]{subfig}
\usepackage{dcolumn}
\usepackage{bm}
\usepackage[hidelinks]{hyperref}
\usepackage{tikz}
\usepackage{siunitx}
\usepackage{amsmath}

\newcommand{\panellabel}[2]{
  \node[anchor=north west, inner sep=1pt, font=\bfseries\small,
        fill=white, fill opacity=0.75, text opacity=1] at (#1) {#2};}

\newcommand{\orcid}[1]{\href{https://orcid.org/#1}
{\includegraphics[width=7pt,height=7pt]{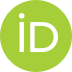}}}

\begin{document}

\preprint{APS/123-QED}

\title{Optomechanical inertial reference for atom interferometry}

\author{Ashwin Rajagopalan\,\orcid{0000-0002-2130-5589}}
\affiliation{Leibniz Universit{\"a}t Hannover, Institut f{\"u}r Quantenoptik, \\Welfengarten 1, 30167 Hannover, Germany}
\author{Daida Thomas\,\orcid{0009-0003-8974-9404}}
\affiliation{Leibniz Universit{\"a}t Hannover, Institut f{\"u}r Quantenoptik, \\Welfengarten 1, 30167 Hannover, Germany}
\author{Alexander Herbst\,\orcid{0000-0003-2351-2554}}
\affiliation{Leibniz Universit{\"a}t Hannover, Institut f{\"u}r Quantenoptik, \\Welfengarten 1, 30167 Hannover, Germany}
\author{Sven Abend\,\orcid{0000-0001-9539-3780}}
\affiliation{Leibniz Universit{\"a}t Hannover, Institut f{\"u}r Quantenoptik, \\Welfengarten 1, 30167 Hannover, Germany}
\author{Ernst M. Rasel\,\orcid{0000-0001-7861-8829}}
\affiliation{Leibniz Universit{\"a}t Hannover, Institut f{\"u}r Quantenoptik, \\Welfengarten 1, 30167 Hannover, Germany}
\author{Knut Stolzenberg\,\orcid{0009-0009-2217-9420}}
\affiliation{Leibniz Universit{\"a}t Hannover, Institut f{\"u}r Quantenoptik, \\Welfengarten 1, 30167 Hannover, Germany}

\author{Felipe Guzm{\'a}n\,\orcid{0000-0001-9136-929X}}
\affiliation{Wyant College of Optical Sciences, The University of Arizona,\\ Tucson, AZ 85721, USA}
\author{Dennis Schlippert\,\orcid{0000-0003-2168-1776}}
\email{schlippert@iqo.uni-hannover.de}
\affiliation{Leibniz Universit{\"a}t Hannover, Institut f{\"u}r Quantenoptik, \\Welfengarten 1, 30167 Hannover, Germany}

\date{\today}

\begin{abstract}
Atom interferometers are among the most sensitive inertial sensors, yet deployment outside the laboratory is limited by vibration noise, conventionally mitigated by external sensors or bulky isolation.
Here we demonstrate a hybrid inertial sensor which fuses an optomechanical resonator and an atom interferometer by exploiting the resonator's test mass as the interferometer's reference mirror.
This allows for ideal correlation as the conversion from mechanical vibrations to optical lattice phase noise is governed by a static and well defined mechanical transfer function, unlike when using commercial accelerometers.
The resonator achieves a displacement sensitivity of \SI{8.6e-15}{m/\sqrt{Hz}} which corresponds to an acceleration sensitivity of \SI{1.1e-6}{ms^{-2}/\sqrt{Hz}} with suppressed $1/f$ noise, demonstrating a high sensitivity measurement bandwidth extending from sub-Hertz to \SI{2.5}{kHz}.
Under ambient laboratory conditions the integrated system removes vibration-induced phase ambiguity for accelerations up to $50$ mm$\,\mathrm{s}^{-2}$ and reaches the interferometer's technical noise limit, which a commercial force-balance accelerometer does not.
Because the cavity resonance-tracking readout is largely independent of the mechanical design, the architecture transfers directly to other precision sensing platforms.
\end{abstract}

\maketitle

\section{\label{sec:Intro}Introduction}

Rapid progress in quantum technologies has driven the transition from laboratory demonstrations to practical applications, with several quantum sensing platforms already achieving performance beyond that of their classical counterparts in specific use cases~\cite{Bongs2019, Pelucchi2022}.
Across the pillars of quantum technologies, mechanical vibrations deteriorate performance:
They lead to qubit dephasing in quantum computing~\cite{PRXQuantum.3.030314, Kono2024}, they induce parametric heating in cavity QED experiments~\cite{PhysRevLett.134.013403, doi:10.1126/sciadv.ads8171}, and they limit the frequency stability in atomic clocks~\cite{Pedrozo-Penafiel2020, Hilton2025,thorpe_measurement_2010} and sensitivity in optical interferometry~\cite{Abadie2011}.
In light-pulse atom interferometers~\cite{kasevich_atomic_1991} accelerations of the inertial reference enter the interferometer phase~\cite{Cheinet_IEEE} and consequently state-of-the-art absolute atom interferometers are usually limited by ambient vibrations.
In dynamic environments, e.g., on moving platforms, parasitic motion causes reduced precision through vibration-induced phase noise and, for large vibration amplitudes, loss of phase information: 
the sinusoidal interferometer output maps the accumulated phase onto the atomic population only modulo $2\pi$, so that excursions beyond a single fringe can no longer be uniquely inverted~\cite{doi:10.1126/sciadv.add3854, Williams2024, Pelluet2025, dArmagnacDeCastanet2024, Rodzinka2024, Stray2022}.
Commonly utilised seismic isolation systems are complex and bulky, impractical for measurements in the field, and their large footprints impair transportability, operation in dynamic environments, and the scalability of quantum sensors~\cite{Panda2024, Robinson2024, Guo2021, Kollár_2015, ATriumphOfSensitivity2017, Kono2024}.
In turn, matter-wave interferometers have successfully been correlated with classical motion sensors to extend the linear range and for operation on noisy, dynamic platforms~\cite{le_gouet_limits_2008,merlet_operating_2009,geiger2011detecting,lautier_hybridizing_2014,Bidel2018}.
Conceptually, the motion sensor tracks the inertial reference mirror motion and allows for real-time or post-correction of vibrations.
However, the achievable vibration correction is fundamentally limited by the spatial separation between the inertial reference and the auxiliary motion sensor and resulting lack of direct coupling.
Typically, the atom interferometer retro-reflection mirror is rigidly attached to the housing of a classical accelerometer rather than to the sensor's test mass.
Consequently, the transfer function between the measured acceleration and the interferometer phase is unknown and potentially time-varying, introducing frequency-dependent errors and cross-coupling between mechanical degrees of freedom.
Realising such an architecture requires optical access to the inertial test mass, which is commonly unavailable in commercial accelerometers and seismometers.
Retro-reflection of the atom interferometer's light field off the motion sensor's test mass has been demonstrated in an electrostatic accelerometer for space applications with limited bandwidth~\cite{zahzam_hybrid_2022}.

\begin{figure*}[t]
\begin{center}
{\includegraphics[width=0.99\textwidth]{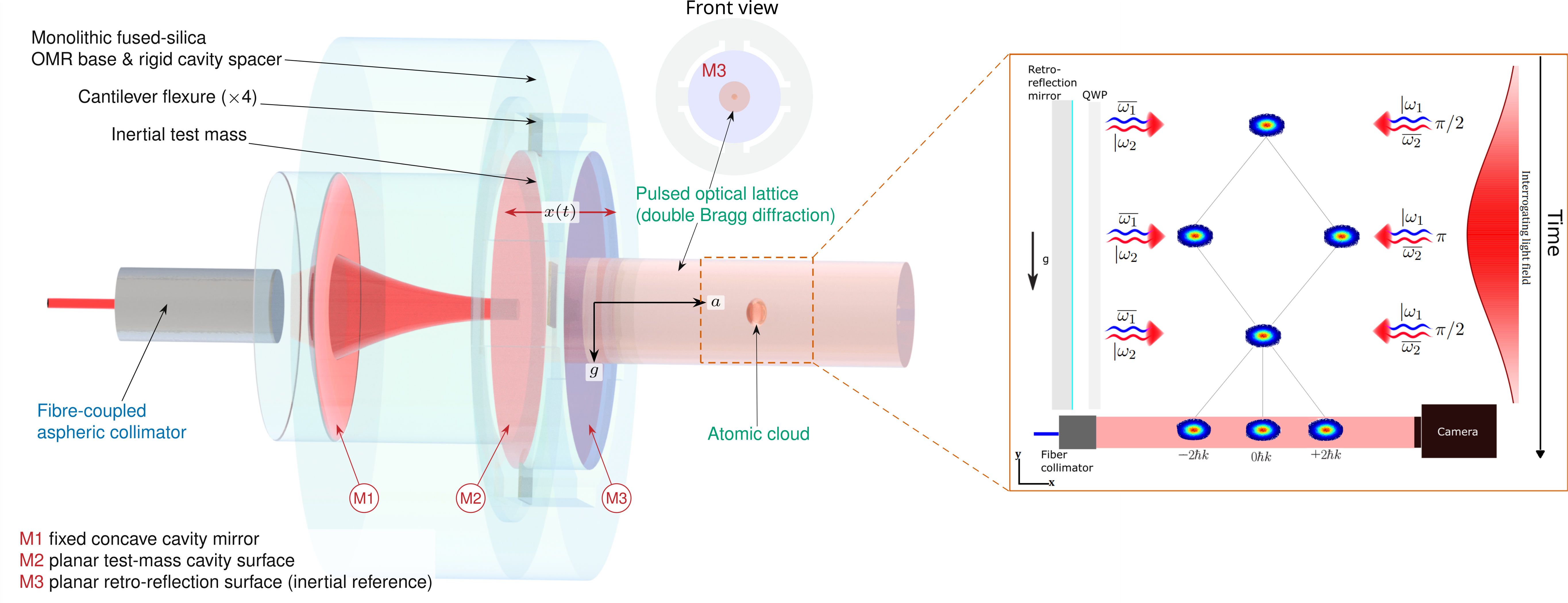}}
\caption{Illustration of the hybrid architecture comprising an atom interferometer, an optical resonator, and an inertial test mass with displacement $x(t)$. 
The zoomed region shows a space-time diagram of a light-pulse atom interferometer employing double-Bragg diffraction with its acceleration measurement axis perpendicular to gravity.}
\label{OMR+AI}
\end{center}
\end{figure*}

Optomechanical systems combine precision optical readout with engineered mechanical resonators, enabling displacement measurements with exceptional sensitivity in compact and scalable devices.
This combination has led to applications in electromagnetic sensing~\cite{Ma2024, Hu2024}, inertial sensing~\cite{Wu2025, Krause2012, Jiao2024, Guzman2014, Zhou2021, Li2024} and the life sciences~\cite{doi:10.1126/sciadv.abq2502, Sansa2020, Yu2016}, and has become a foundational platform for quantum technologies~\cite{Xia2023, Huang2024, Yu2022, PhysRevLett.130.263603, Barzanjeh2022, Verhagen2012}.
These capabilities make optomechanical resonators a natural platform for realising inertial references in atom interferometry.
Compared with state-of-the-art commercial accelerometers, they sustain a comparably flat noise floor across a bandwidth several times wider and correlate with the interferometer through a static, well-defined mechanical transfer function.
Here we fuse the passive retro-reflection mirror of an atom interferometer with a high-sensitivity optomechanical resonator (OMR) that simultaneously serves as the interferometer's inertial reference~\cite{richardson_quantum_2020} and as a vibration sensor.
The result of our integration is that vibration correction becomes limited by the atom interferometer itself rather than by the reference: 
under ambient vibration of \SI{50e-3}{ms^{-2}} the corrected interferometer signal reaches its technical noise limit of $\sigma_{\mathrm{OMR}}=0.051$, whereas simultaneous correction using a commercial force-balance accelerometer remains limited to $\sigma_{\mathrm{acc}}=0.069$.
To our knowledge this is the first demonstration in which an integrated optical reference matches, and under strong vibration surpasses, a broadband commercial seismic accelerometer in correcting an atom interferometer -- while being the retro-reflecting surface that defines the interferometer phase.

\section{Results}

\subsection{Apparatus}
Our hybrid sensor, illustrated in Fig.~\ref{OMR+AI}, comprises a mechanical drum-head resonator, a monolithic test mass suspended on cantilever flexures, that mediates direct coupling between the atom interferometer phase and continuous displacement readout provided by a high-finesse optical cavity and a double-Bragg atom interferometer~\cite{PhysRevLett.116.173601}.
The integration of an optomechanical resonator replaces the unknown, potentially time-varying transfer function between the inertial reference and the auxiliary vibration sensor -- the factor that limits conventional hybrid atom interferometers -- with the single static, well-characterised response of one mechanical element.
As a classical benchmark we operate a commercial force-balance accelerometer (Nanometrics Titan) alongside the resonator, an instrument that has been used successfully to post-correct atom interferometers~\cite{Pelluet2025, Menoret2018}.
It is mounted beside the OMR mount on the same optical breadboard but approximately \SI{10}{cm} off the Bragg optical axis and records the ambient acceleration simultaneously with every interferometer shot.

\subsection{Characterisation of the OMR}
\begin{figure}[t]
\centering
\begin{tikzpicture}
  \node[anchor=south west,inner sep=0] (img) at (0,0)
    {\includegraphics[width=0.5\textwidth]{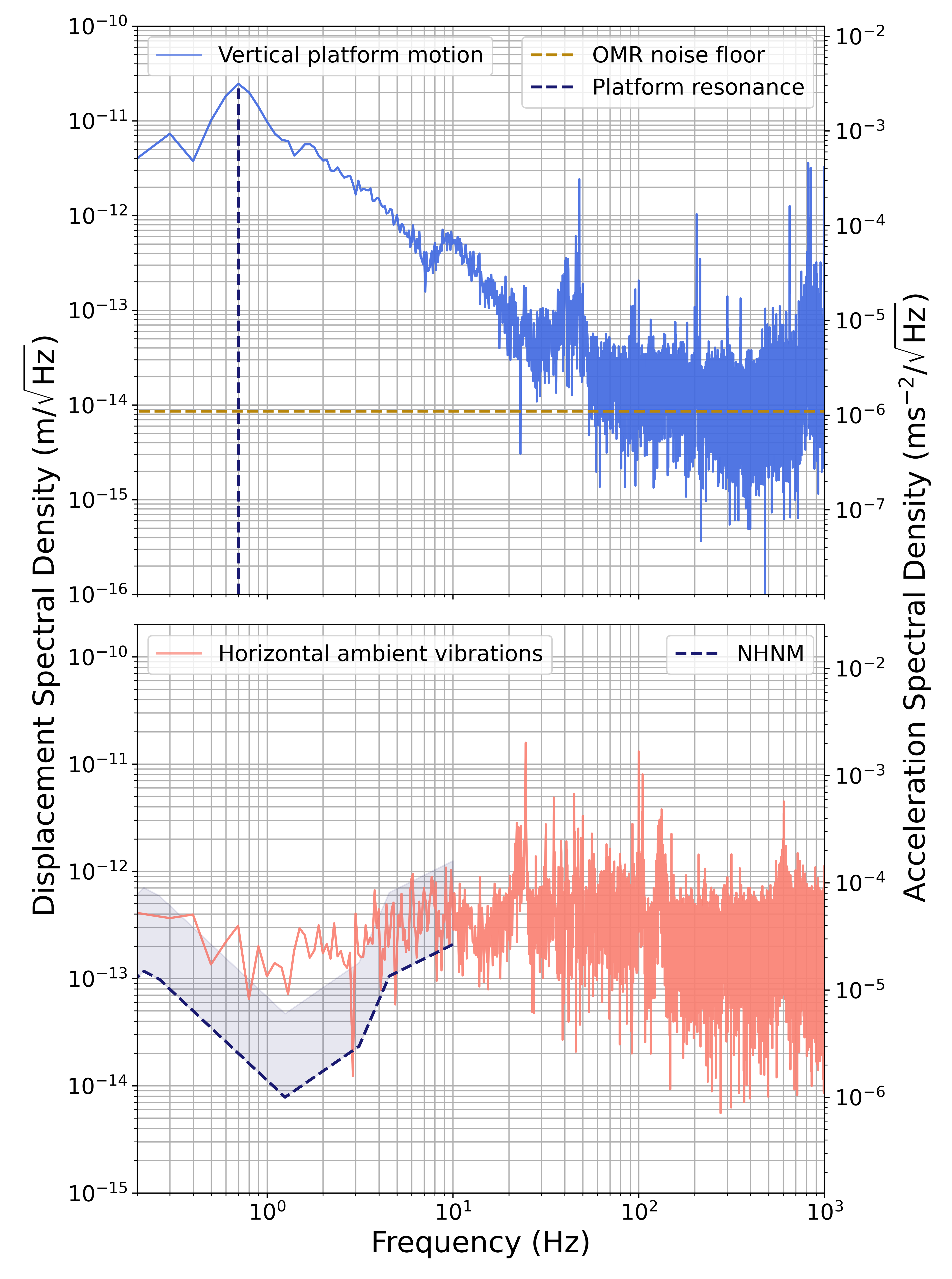}};
  \begin{scope}[x={(img.south east)},y={(img.north west)}]
    \panellabel{.15,0.575}{(a)}
    \panellabel{.15,0.11}{(b)}
  \end{scope}
\end{tikzpicture}
\caption{(a) Measuring vibration isolation platform motion oscillating at its \SI{700}{mHz} natural frequency using the optomechanical resonator. The suppressed ambient vibration environment enables reaching its intrinsic sensitivity limit which is set by the photo detector shot noise at frequencies $>\SI{50}{Hz}$. (b) Ambient vibration noise measurement with the same coupling configuration to the atom interferometer without the use of vibration isolation. The white noise floor is limited by ambient vibrations and at lower frequencies the spectral density shows a characteristic NHNM dip.}
\label{OMR_sensitivity}
\end{figure}

The optomechanical resonator measures the displacement of its mechanical test mass relative to its rigidly coupled base, from which acceleration is inferred using the driven harmonic oscillator transfer function (cf. Methods).
Building on previous work~\cite{Richardson}, displacement is measured by a high-finesse optical cavity formed between a fixed concave mirror (M1) and the mechanical test mass, which is also the flat mirror (M2) of a plano-concave Fabry-Perot cavity, as depicted in Fig.~\ref{OMR+AI}.
The resonator's fundamental angular resonance frequency is determined at $\omega_o= 2\pi\cdot\SI{1800}{Hz}$, far above the atom interferometer's bandwidth, with a mechanical quality factor of 33 under ambient atmospheric pressure.
Placing the resonance above the interferometer band ensures that the mechanical response is flat where the interferometer is sensitive, while the choice of resonance frequency simultaneously sets the trade-off between acceleration sensitivity and dynamic range.
The cavity exhibits a free spectral range of \SI{8.69}{GHz} and a measured intrinsic linewidth of \SI{2.79}{MHz} corresponding to a finesse $\mathcal{F}=3118$ at its fundamental TEM$_{00}$ mode, consistent with the value targeted by the \SI{99.9}{\percent} coating specification and confirming that the engineered optical design was realised.
Low-frequency optical sensing is usually limited by 1/f noise processes~\cite{Jin2021}.
Our readout scheme uses lock-in detection via modulating the probe beam frequency and optical resonance tracking via a feedback loop, this substantially reduces 1/f noise and grants access to signals in the sub-Hertz regime while increasing dynamic range.
Sensor operation on a commercial vibration isolation platform allows the intrinsic noise floor of the optomechanical resonator to be resolved for frequencies \SI{>50}{Hz} where performance is limited by photon shot noise owing to the very low optical power of $\approx \SI{2}{\micro W}$ reflected back from the cavity, closely followed by relative intensity noise (Fig.~\ref{OMR_sensitivity}a).
We infer a displacement noise floor of \SI{8.6e-15}{m/\sqrt{Hz}} which corresponds to an acceleration noise floor of \SI{1.1e-6}{ms^{-2}/\sqrt{Hz}} up to $\sqrt{2}\cdot\omega_0=\SI{2.5}{kHz}$.
Together with the maximum acceleration resolved during the platform measurement, this noise floor corresponds to a measured dynamic range of \SI{125}{dB}; the range accessible in principle, bounded by the tracking bandwidth of the electro-optic modulator and the mechanical stress limit of the test mass, extends to \SI{161}{dB}.

To validate the inertial response, the resonant motion of the vibration isolation platform at its natural frequency of \SI{700}{mHz} was measured simultaneously using the optomechanical resonator and the commercial force-balance accelerometer.
In Fig.~\ref{OMR_sensitivity}b, recorded in horizontal orientation and without vibration isolation, ambient vibrations dominate the measured spectrum, while the resonance of the isolation platform is absent.
Both the optomechanical resonator and the commercial accelerometer exhibit the characteristic dip around \SI{1}{Hz} predicted by Peterson's New High Noise Model (NHNM)~\cite{peterson,nhnm1,nhnm2}, a standard reference spectrum for ambient seismic background, confirming accurate low-frequency inertial measurements under ambient conditions.

\begin{figure*}[tb]
	\centering
	\includegraphics[width=0.95\textwidth]{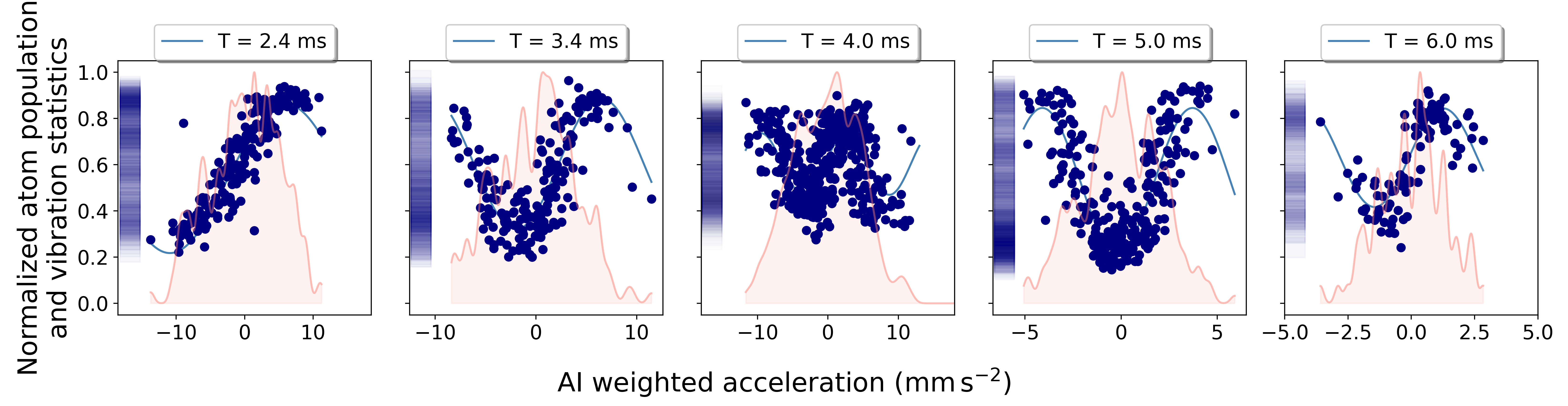}
	\caption{Ambient vibration noise post-correction using the optomechanical resonator signal for pulse separation times $T=\SIrange{2.4}{6}{ms}$. The scattered data points are the normalised atomic population with removed phase ambiguity via vibration noise post-correction comparison to the expected responses (solid blue curves), the gradient density bar highlights population distribution indicating atom interferometer technical noise. The normalised distribution of atom interferometer weighted vibration noise which randomly scans over its phase is depicted in the background (red curve).
 }

	\label{PostCorrection}
\end{figure*}
\subsection{Sensor Fusion with an Atom Interferometer}
Utilising a standard mirror for the atom interferometer instead of the optomechanical resonator in a test measurement campaign does not add detectable excess noise and the yielded standard deviations $\sigma_{\mathrm{OMR}} = \num{0.0855(70)}$ and $\sigma_{\mathrm{mirror}} = \num{0.0844(69)}$, consistent within uncertainty.
This is expected: the interferometer's corner frequency of $1/2T$ reaches at most \SI{250}{Hz} for the shortest pulse separation used here, well below the fundamental mechanical resonance at \SI{1800}{Hz}, and the vibration-induced phase falls off steeply above the corner frequency~\cite{Cheinet_IEEE}.
We utilise the optomechanical sensor to restore atom interferometer performance under ambient vibration. 
Figure~\ref{PostCorrection} shows vibration post-correction for atom interferometers operated with pulse separation times from $T=\SIrange{2.4}{6}{ms}$.
For every experimental shot, the vibration-induced atom interferometer phase is reconstructed from the simultaneously recorded optomechanical resonator signal using the atom interferometer sensitivity function~\cite{Cheinet_IEEE,AI_sens_form_PhysRevA.92.023626} as described in the Methods section.
The reconstructed interference signal is shown as blue data points, while the corresponding normalised atom population distribution is represented by the blue density bars.
The expected interferometer response is indicated by the solid blue curve.
The weighted acceleration noise, obtained by convolving the measured acceleration with the atom interferometer sensitivity function, is shown for each data set.
The reconstructed vibration phase successfully recovers the interferometer fringe across all investigated pulse separation times.
For pulse separation times up to $T=\SI{6}{ms}$, the recovered interference signal is consistent with the intrinsic technical noise of the atom interferometer.
As an independent validation, post-correction using a commercial force-balance accelerometer yields comparable performance under these vibration conditions.
That the external sensor also reaches the interferometer's technical noise limit is itself informative: had the resonator introduced motion inaccessible to a separate sensor, this agreement could not have been obtained.
Under substantially stronger vibration, however, the two references diverge, motivating a direct comparison.

\begin{figure*}[!tb]
	\centering
 \subfloat[]{\includegraphics[width=0.495\textwidth]{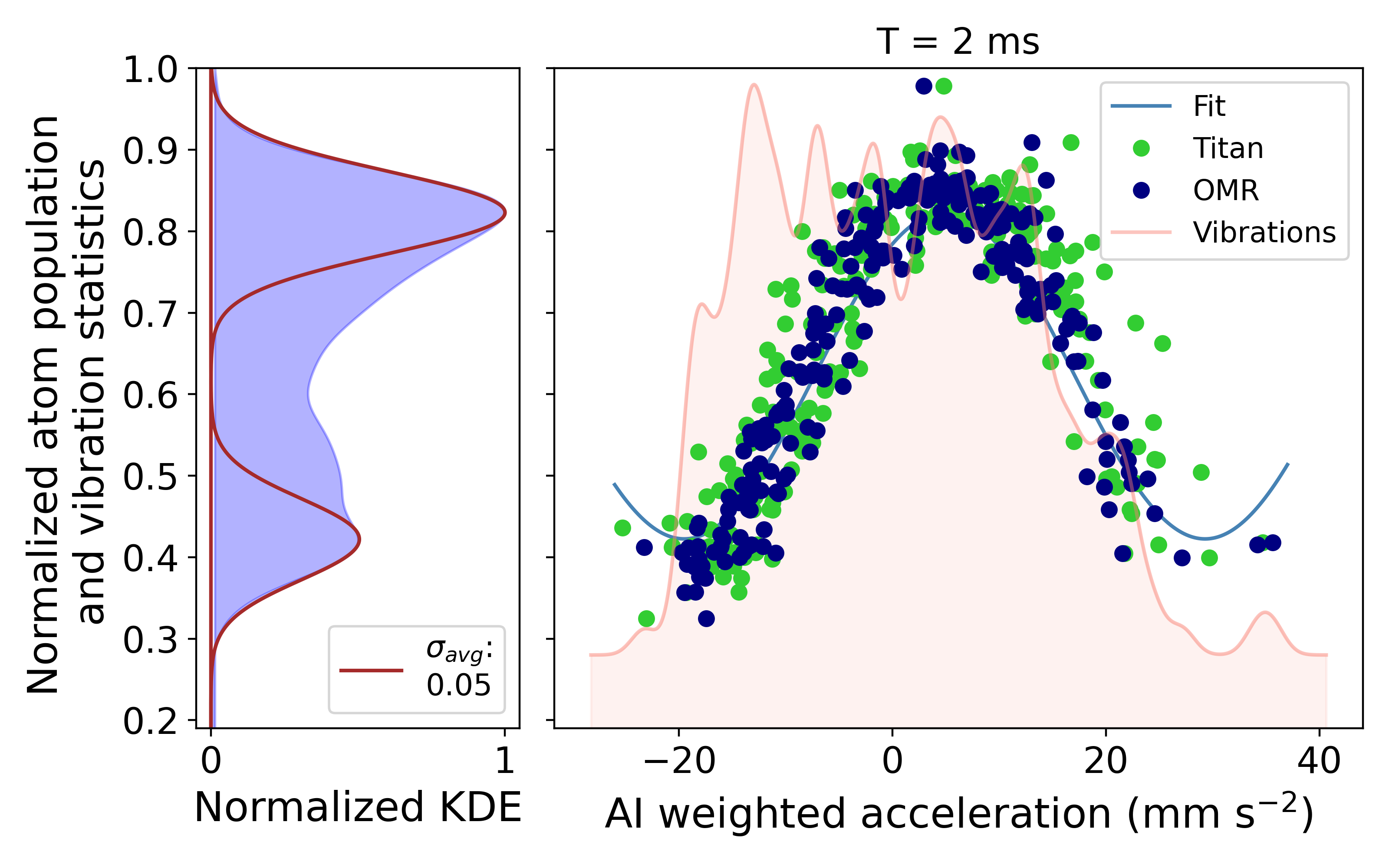}\label{OMRvsTitan_Fr}}
 \hfill
 \subfloat[]{\includegraphics[width=0.495\textwidth]{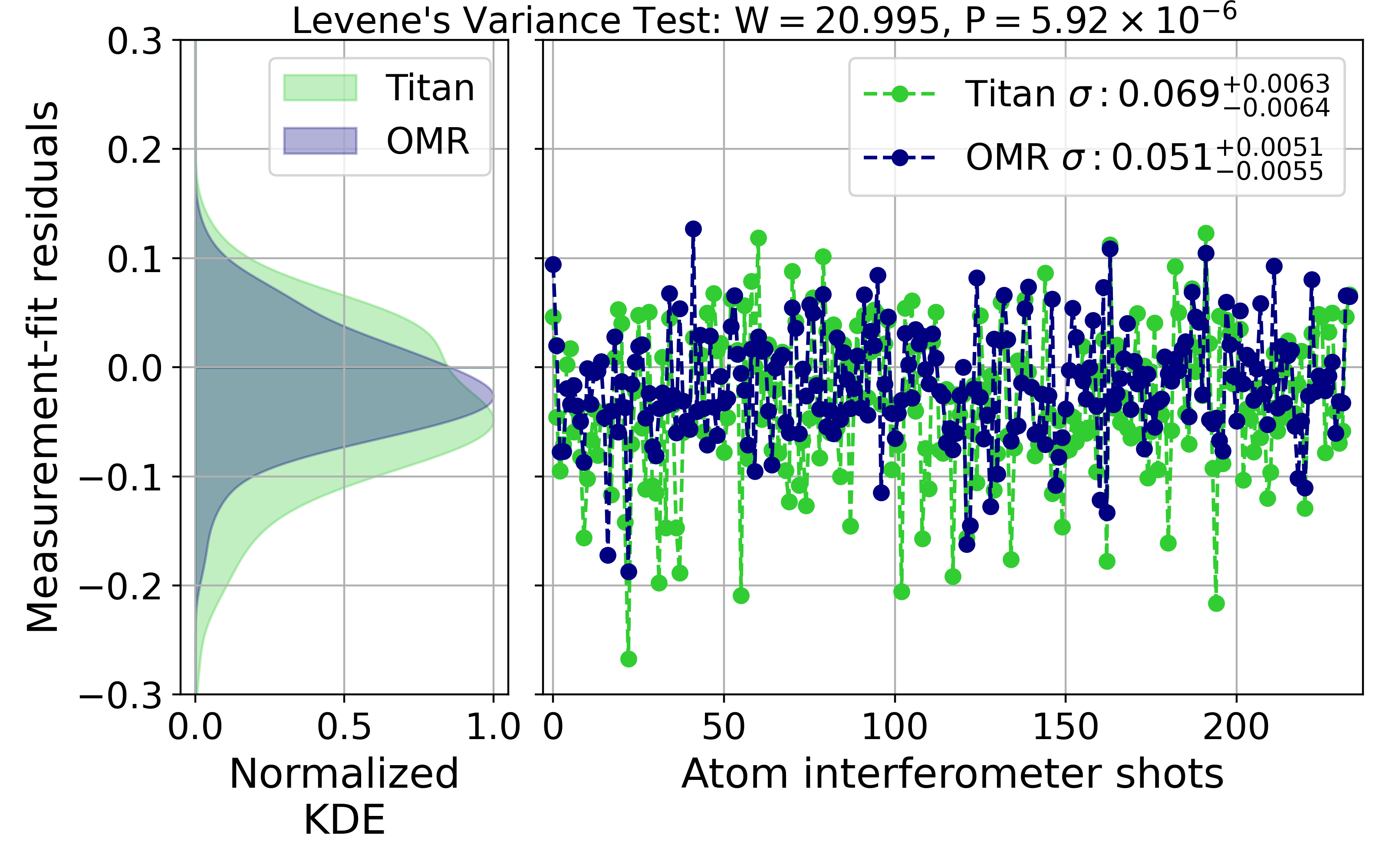}\label{OMRvsTitan_res}}
	\caption{(a) Post-corrected $T=\SI{2}{ms}$ atom interferometer operated under very high ambient vibration correlated with the optomechanical resonator (blue data points) and a simultaneous comparison to the commercial accelerometer (green data points). 
    The solid blue curve displays the expected atom interferometer response fitted to the OMR data. 
    Coupled vibration noise induces an overall spread of $\approx50$ mm/s$^2$ on the atom interferometer phase response. The atom interferometer technical noise is estimated by Gaussian fitting the individual peaks of its bi-modal kernel density estimation, amounting to an average value of $\sigma_{\mathrm{avg}}=$ 0.05. 
    (b) Vibration noise suppression performance using the Titan and optomechanical resonator is evaluated by independently calculating their residuals between the expected atom interferometer response and the post-corrected. 
    Under such high ambient vibrations the more direct coupling of the OMR allows us to reach the atom interferometer's technical noise limit while the reference accelerometer does not.}
	\label{OMRvsTitan}
\end{figure*}

We therefore evaluate correlated vibration noise post-correction for a $T=\SI{2}{ms}$ atom interferometer using the optomechanical resonator and the classical accelerometer signal, shown in Fig.~\ref{OMRvsTitan_Fr}, where very high ambient vibrations of around $50$ mm$\,\mathrm{s}^{-2}$ scan across the interferometer response.
This campaign coincided with pronounced anthropogenic activity in the laboratory and therefore emulates high-noise conditions as expected outside of the laboratory.
The atom interferometer technical noise is estimated by Gaussian least squares regression of its bi-modal distribution, the $\sigma_{\mathrm{avg}}$ of the two peaks is 0.05.
The correlation of large ambient vibrations using the classical accelerometer (green data points) is noisier than the one of the optomechanical resonator (blue data points) as compared to the expected response in Fig.~\ref{OMRvsTitan_Fr} shown by the solid blue curve.
This finding is backed by calculating the difference between the recovered signals and the expected response as shown in Fig.~\ref{OMRvsTitan_res}.
The standard deviation $\sigma$ of the residuals using the two accelerometers show that vibration induced phase correction using the optomechanical resonator reaches the atom interferometer technical noise of 0.05, whereas the Titan correction does not.
With 234 correlated shots in each data set, we quantified the uncertainties using percentile bootstrapping, a method robust to non-normal distributions that provides strict upper and lower limits. 
This yielded a \SI{95}{\percent} confidence interval of $\sigma_{\mathrm{acc}}\,=[0.0626, 0.0753]$ for the Titan correction and $\sigma_{\mathrm{OMR}}\,=[0.0455, 0.0561]$ for the OMR correction. 
To evaluate the statistical significance of this difference, we applied Levene's test.
The test returned a statistic of $W = 21$ and a p-value of $p = 5.92 \times 10^{-6}$. 
This low probability confirms that the disparity in variance between the two corrections is highly significant, rather than a mere statistical fluctuation of a single campaign. 
That the campaign was subject to genuinely elevated external vibration, rather than to motion peculiar to the resonator, is independently confirmed by the ambient acceleration recorded simultaneously by the reference accelerometer.

Because the reference mirror and inertial sensor are integrated within the same physical object, the integrated architecture enables vibration correction to the atom interferometer's technical noise limit even under large ambient vibration amplitudes, whereas correction based on a spatially separated classical accelerometer remains limited by imperfect correlation.

\section{Discussion}

In conventional hybrid atom interferometers, the auxiliary motion sensor is mounted in proximity to, but spatially separated from, the inertial reference.
Consequently, vibration correlation relies on an unknown and potentially time-varying mechanical transfer function between the two sensors.
Under dynamic motion this transfer function degrades common-mode rejection through sensor misalignment and cross-coupling of additional degrees of freedom, leading to imperfect recovery of the vibration-induced interferometer phase~\cite{Bidel2018}.
The broader implications of unknown transfer functions are discussed in Ref.~\cite{bronstein2024uncovering}, while their impact on hybrid atom interferometers has previously been identified in Refs.~\cite{geiger2011detecting,Bidel2018}.
In order to compensate for non-ideal correlation, sophisticated algorithms yielding only small improvements~\cite{Hyb_10.1116/5.0304770} and algorithms that lower the pulse separation time T~\cite{Bidel2018} to suppress this effect have been implemented.
Rather than compensating for an unknown transfer function, our architecture simplifies it by merging the inertial reference and auxiliary vibration sensor within the same mechanical element.
The same argument applies wherever a retro-reflected optical lattice or a cavity mirror defines an optical phase reference. 
In each case the quantity that must be known is the motion of one specific optical surface, and an intergated motion sensor reports it directly. 
To our knowledge, no other inertial reference has been demonstrated that both reports the motion of the phase-defining optical surface at that surface itself and does so across a broad band; 
the electrostatic implementation of Ref.~\cite{zahzam_hybrid_2022} achieves direct coupling to the accelerometer test mass, but over a restricted bandwidth designed for space-borne Earth observation.

The optomechanical accelerometer's resonance frequency, and with it the trade-off between acceleration sensitivity, bandwidth and dynamic range, can be chosen for a given application without altering the readout. 
Where the optical surface must itself remain rigid, the sensor can instead be bonded to the rear of a conventional optic. 
Integration within the same mechanical element is then no longer exact, but the residual mechanical path is short, stiff and static, in contrast to the extended and poorly characterised coupling between an interferometer mirror and a separate accelerometer.

The fully vacuum-compatible, non-magnetic glass design enables direct integration into compact atom interferometers or other quantum technology platforms without additional inertial sensing hardware or electromagnetic interference.
The resonator can therefore be incorporated into deployable quantum sensors, such as compact quantum gravimeters~\cite{Abend_PhysRevLett.117.203003}, while leveraging the existing optical infrastructure required for atom interferometry.

Placed alongside the wider field of optomechanical inertial sensing, the resonator is competitive rather than exceptional in displacement sensitivity: 
comparable or lower displacement noise floors have been reported for cavity- and membrane-based accelerometers, several of them packaged and operable outside vacuum~\cite{Guzman2014, Zhou2021, Li2024, Jiao2024, Wu2025}.
Displacement sensitivity alone, however, is not the relevant figure of merit for an inertial reference.
These are excellent standalone accelerometers, and the distinction drawn here is one of function rather than of sensitivity: 
none of them can serve as the retro-reflecting surface that defines the atom interferometer phase.
For the combined sensor the consequence is that the accuracy of the vibration correction is bounded by the technical noise of the atom interferometer itself, and no longer by the fidelity of a correlation between two separate objects.
The noise-equivalent acceleration these devices report is moreover achieved within a kilohertz-scale band and degrades towards lower frequencies, where 1/f noise processes in the optical readout dominate, whereas an atom interferometer responds significantly only below a corner frequency of $1/2T$, at most \SI{250}{Hz} for T = 2 ms; 
the resonance-tracking lock-in readout suppresses 1/f noise and enables enhanced sensitivity at lower frequencies that is relevant for vibration noise correlation.
Because this readout addresses the optical cavity rather than the mechanical design, it is largely independent of resonator geometry and is directly transferable to other resonator designs and sensing platforms.

Several limitations remain. 
The displacement noise floor is currently set by photon shot noise owing to the $\approx \SI{2}{\micro W}$ recovered from the cavity. 
Improving the optical coupling efficiency, and with it the returned power, would reduce the shot-noise contribution as the inverse square root of the power. 
Relative intensity noise scales with the power and would then become the limiting term, requiring active intensity stabilisation of the source.
The demonstration is confined to a single sensitive axis and to a static laboratory environment; 
operation on a moving platform, where a static transfer function should be most advantageous, has yet to be shown. 
The pulse separation times accessible in this apparatus are short, bounding the interferometer's intrinsic acceleration sensitivity, and the advantage over an external accelerometer is established in the high-vibration regime rather than across all operating conditions. 
Sensitivity and dynamic range also remain coupled through the choice of resonance frequency and cavity finesse.
Post-correction is only one way to use this signal.
The displacement readout is continuous with kilohertz bandwidth, so the same signal could in principle drive real-time actuation~\cite{lautier_hybridizing_2014,cheiney_navigation-compatible_2018} to stabilise the optical phase rather than correcting it afterwards. 
This has not been demonstrated here.

Future improvements, including cyclic de-biasing~\cite{cheiney_navigation-compatible_2018} and photonic integration~\cite{isichenko_photonic_2023}, could enable long-term stable hybrid quantum inertial sensors for resilient navigation.
Beyond atom interferometry, the presented integrated hybrid sensing architecture provides a general framework for integrating precision optical displacement sensing directly into quantum inertial sensors, opening new opportunities for compact and field-deployable quantum technologies.

\section{Methods}

\begin{figure*}
	\centering
 \subfloat[]{\includegraphics[width=0.31\textwidth]{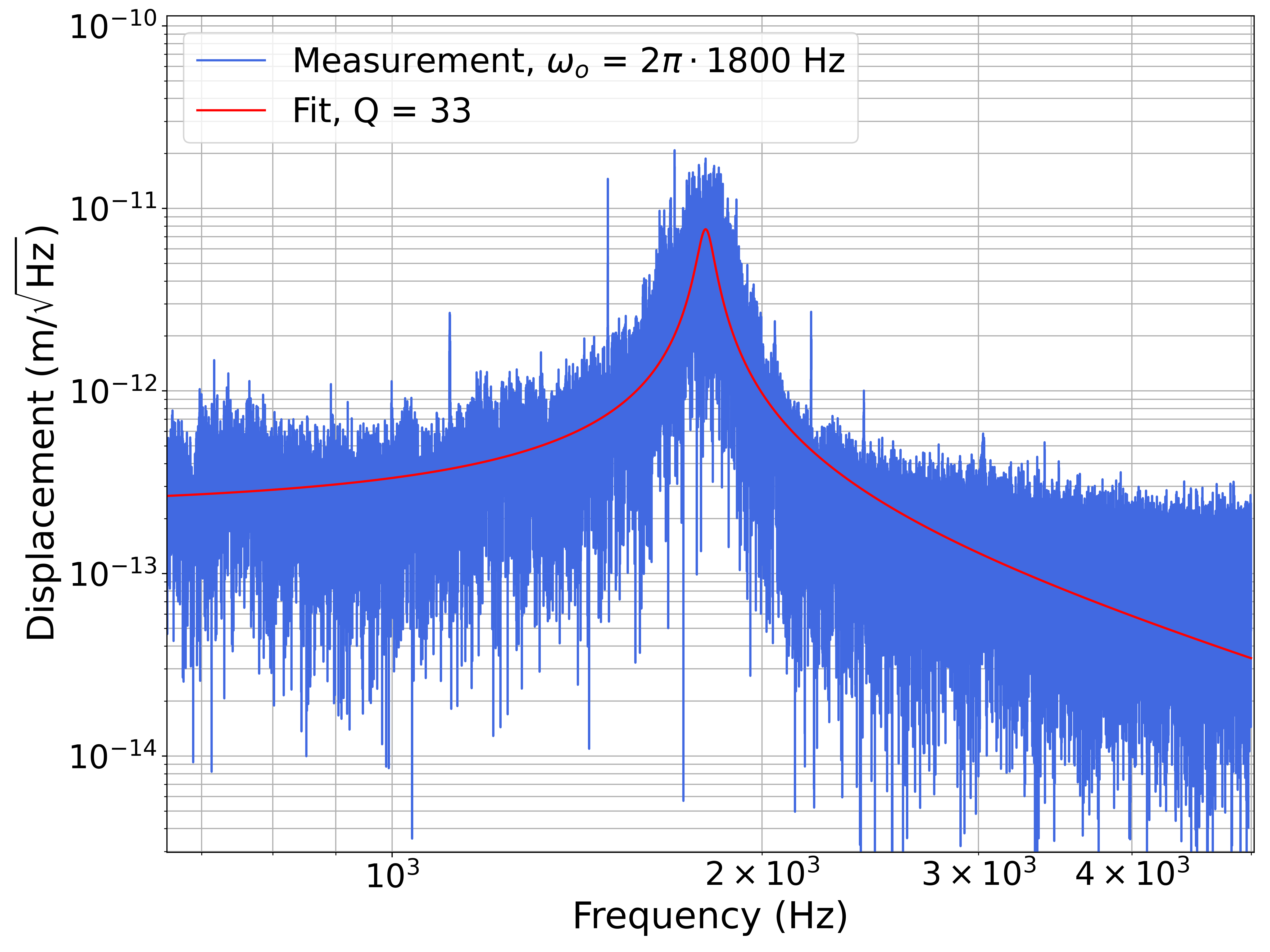}\label{OMRchar_mech}}
	\hfill
 \subfloat[]{\includegraphics[width=0.31\textwidth]{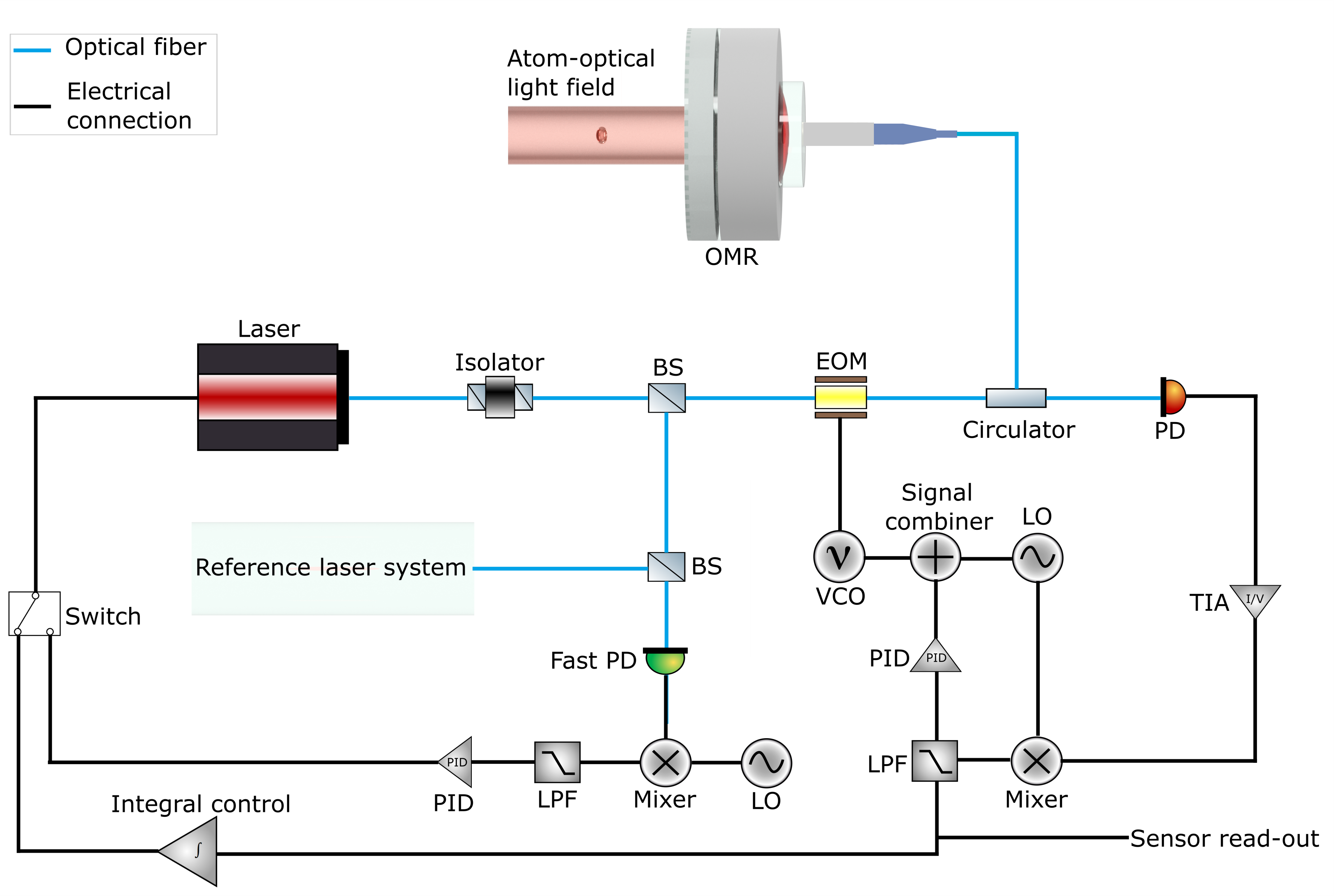}\label{exp_readout}}
 \hfill
 \subfloat[]{\includegraphics[width=0.31\textwidth]{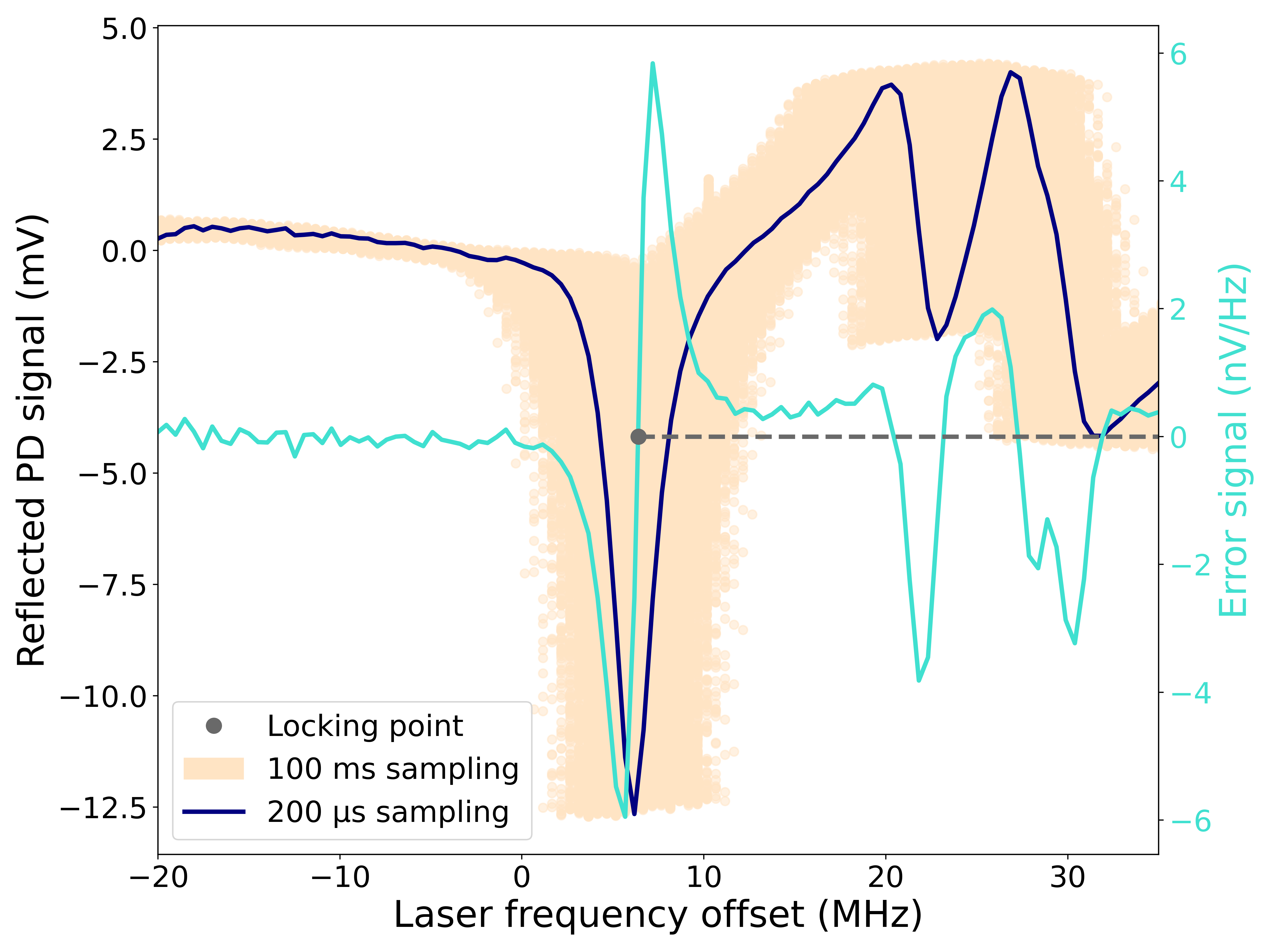}\label{error_sig}}
	\caption{(a) Measured fundamental in-plane mode of the mechanical resonator via optical readout for operation under ambient atmosphere. (b) The complete optical readout schematic for measuring using the optomechanical resonator. It is a lock-in detection feedback loop based readout scheme using a single EOM sideband that efficiently suppresses low frequency readout noise and increases dynamic range despite the small optical resonator linewidth. The DC source laser optical frequency can possibly be stabilised to an external stable reference or to the probing optical cavity frequency as per experimental convenience to perform vibration measurements, this is indicated by a switch in the schematic. (c) The measured intrinsic Gaussian linewidth of \SI{2.79}{MHz} of the probed optical mode and the generated error signal via lock-in detection are shown. The error signal serves as a frequency discriminator converting vibration induced cavity length fluctuations into a measurable voltage, this is used for readout and to lock the single EOM sideband to the probing cavity mode using a feedback loop. This setup tracks center frequency variations of the cavity mode due to cavity length fluctuations, larger sampling durations corresponding to \SI{100}{ms} (\SI{10}{Hz}) shows significant coupling of ambient vibrations whereas shorter timescales of \SI{200}{\micro s} (\SI{5}{kHz}) exposes its intrinsic linewidth as it is well above the natural frequency of \SI{1.8}{kHz} in its vibration isolation regime.}
    \hfill
	\label{signal}
\end{figure*}
\subsection{Atom interferometer}

A \textsuperscript{87}Rb atom interferometer employing double Bragg diffraction with an acceleration measurement axis that is orthogonal to gravity as depicted in Fig.~\ref{OMR+AI}, is used for hybridisation with the optomechanical resonator.
The pulse separation time is constrained by the geometry of this apparatus: 
because the interferometer axis is orthogonal to gravity, the atoms fall through the Bragg beam during interrogation, so the available interferometry baseline is set by the beam diameter of \SI{8}{mm} ($1/e^2$). 
Longer pulse separation times are therefore a limitation of the present apparatus rather than of the optomechanical inertial reference, which is agnostic to $T$. 
A comprehensive description of this atom interferometer experiment is presented in reference~\cite{Knut_PhysRevLett.134.143601}.
The leading order phase of the atom interferometer response is given as,
\begin{equation}
 \Delta \phi \,=\, \mathrm{k_{eff}}T^2a
\end{equation}
where $\mathrm{k_{eff}}$ is the effective wave vector of the light pulses manipulating the atoms, $T$ is the pulse separation time and $a$ is the acting acceleration.
Therefore, the atom interferometer is used as an absolute reference to calibrate the optomechanical resonator via vibration noise correlation.

The sensitivity function $G(\omega)$ is a central concept in quantum sensing and atom interferometry~\cite{Cheinet_IEEE}. 
It describes, in the frequency domain, how sensitive the system is to phase-modulating noise at a specific modulation frequency $\omega$. 
Using this function, the impact of environmental disturbances such as vibrations in this work on the final measurement signal can be precisely quantified. 
For a specific pulse sequence depending on the Rabi frequency $\Omega_R$, the free evolution time $T$, and the pulse duration $\tau_R$ the sensitivity function can be formulated as follows:

\begin{equation}
\begin{split}
G(\omega) &= \frac{4i\Omega_R}{\omega^2 - \Omega_R^2} \sin\left(\frac{\omega(T + 2\tau_R)}{2}\right) \\
&\quad \times \left( \cos\left(\frac{\omega(T + 2\tau_R)}{2}\right) + \frac{\Omega_R}{\omega} \sin\left(\frac{\omega T}{2}\right) \right).
\end{split}
\end{equation}

\subsection{Optomechanical resonator}

The optomechanical resonator presented in this article is specifically designed to serve as the retro-reflecting inertial reference for atom interferometry and is machined from monolithic fused silica. The OMR has an outer diameter of 50.8 mm and a length of 30.117 mm from the planar surface of M1 to M3 including the optical cavity as depicted in Fig.~\ref{OMR+AI}. Rigidly attached to the planar surface of M1 is the aspheric fiber collimator which has a length of 36.8 mm independent of the optical fiber.
Its inertial test mass has a weight of \SI{10.9}{grams}, is suspended with the help of four cantilever flexures, and forms a driven harmonic oscillator following motion governed by,
\begin{equation}
 \frac{X(\omega)}{A(\omega)} = -\frac{1}{\omega_{o}^2-\omega^2+i\left(\frac{\omega_{o}}{Q}\omega\right)}
\end{equation}
where $X(\omega)$ and $A(\omega)$ are the displacement and acceleration frequency responses respectively, $\omega_o$ is the angular resonance frequency of the mechanical resonator and $Q$ is its quality factor. For $\omega << \omega_{o}$ the displacement to acceleration relation follows $X(\omega) \propto -\frac{1}{\omega_{o}^2}A(\omega)$.
A key tuning parameter that governs sensitivity, bandwidth and dynamic range of the OMR is $\omega_o$, for this application it was designed to be around $\omega_o=2\pi\cdot\SI{1800}{Hz}$ as the measurement shows in Fig.~\ref{OMRchar_mech} with a $Q=33$ under ambient atmosphere.
Higher-order mechanical modes lie well above the fundamental by design, as confirmed by finite-element analysis, and are further suppressed by their lower quality factors; they were not resolved in the optical readout. They are therefore irrelevant to the atom interferometer, which has its highest phase sensitivity only below its corner frequency of $1/2T$, at most \SI{250}{Hz} for the shortest pulse separation used here and decreases as $1/f^{4}$ for higher acceleration frequencies. One flat surface of the cylindrical test mass is dielectric coated for \SIrange{720}{850}{nm} with a reflectivity greater than \SI{99.9}{\percent} in order to retro-reflect the light pulses for atom interferometry using $^{87}$Rb and $^{39}$K atoms~\cite{Schlippert_PhysRevLett.112.203002, Herbst2024}.
The other flat surface is dielectric coated to reflect 1560 nm around a wide-band with a reflectivity of \SI{99.9}{\percent}, this along with a fixed curved mirror constitutes a stable fibre coupled high finesse optical resonator with a cavity length of \SI{17.25}{mm} which is used to directly measure mirror motion.

\noindent\textsc{Readout scheme:}\\
The other crucial part that governs overall measurement performance of the optomechanical resonator is its optical readout scheme, this is equally important as possessing good optical and mechanical properties.
The perspective of significantly increasing optical cavity reflectivity compared to reference~\cite{Richardson} is to enhance sensitivity and suppress 1/f noise processes.
Although, increasing the reflectivity substantially decreases dynamic range of the optomechanical resonator due to the sensitivity trade-off.
The lock-in detected feedback loop based readout scheme shown in Fig.~\ref{exp_readout} enables highly sensitive measurements with a large dynamic range.
Symmetric sidebands are generated around the source laser carrier frequency using an Electro-Optical Modulator (EOM), where one of the sidebands is used to probe and track an optical cavity mode.
Reflected light from the cavity is exclusively collected on a Photo Detector (PD) and the photo-current is converted to voltage with a low-noise Trans-Impedance Amplifier (TIA).
The EOM sideband is modulated at hundreds of kilohertz using a Voltage Controlled Oscillator (VCO) which is driven by a Local Oscillator (LO).
The PD signal is mixed down with the LO at the same modulation frequency and is phase detected using a RF mixer plus Low Pass Filter (LPF) combination to produce the error signal that is depicted in Fig.~\ref{error_sig}.
The error signal serves as a frequency discriminator that converts vibration induced optical cavity length into a measurable voltage.
This is used as the sensor readout which is an instantaneous measurement of the vibration induced optical cavity displacement of the OMR.
The error signal is also input to the Proportional-Integral-Derivative (PID) controller which also drives the VCO to continuously track fluctuations in the probed optical cavity mode.
For a faster scan frequency well above the mechanical resonance the intrinsic linewidth is exposed, whereas for slower scans or larger sampling times there is vibration coupled broadening which is being tracked as shown in Fig.~\ref{error_sig}.
Therefore, the LO and PID controller simultaneously drive the VCO with the help of a signal combiner in order to establish the Lock-in detected feedback loop enabled readout scheme.
Lock-in detection by demodulating a high frequency probe signal facilitates isolation of 1/f noise processes due to laser intensity and optical setup noises, and the feedback loop enables large dynamic range operation by means of locking to the cavity resonance center frequency.
There are two possibilities to set the DC optical frequency of the source laser as depicted in Fig.~\ref{exp_readout} with a switch, one is with reference to Rb spectroscopy and the other is by locking to the probed OMR optical cavity mode.
For method one the OMR source laser is frequency offset locked to a reference laser system that is stabilised to the $^{85}$Rb D2 transition via Modulation Transfer Spectroscopy (MTS), an independent LO is used to set the offset frequency from the atomic reference and the beat signal is generated using a large bandwidth fast PD.
For method two the source laser is simply locked to the probed OMR cavity mode via a slow integral controller, this enabled stable feedback loop operation as it does not depend on the LO frequency range.
Both methods were used for hybridisation with the atom interferometer, but only the AC vibration accelerations from the OMR are used to calculate the atom interferometer phases.

\section{Data availability}
The data used in this manuscript are available from the corresponding author upon reasonable request.

\appendix

\bibliography{reference}

\section{Acknowledgements}
Funded by the Deutsche Forschungsgemeinschaft (DFG, German Research Foundation) under Germany's Excellence Strategy – EXC 2123/2 QuantumFrontiers – 390837967 and within CRC 1464 (TerraQ), project A02.
This project is furthermore supported by the German Space Agency at the German Aerospace Center with funds provided by the Federal Ministry for Economic Affairs and Climate Action due to an enactment of the German Bundestag under Grant Nos. DLR 50NA2106 (QGyro+).

\section{Author contributions}
D.S., F.G., and E.M.R. conceived the idea of combining an atom interferometer and optomechanical resonator to harness complementary advantages.
D.S. is the principal investigator of the project.
F.G. provided the preliminary design for the mechanical resonator that serves as the retro-reflection mirror.
E.M.R. contributed to the design and construction of the atom interferometry apparatus.
A.R. finalised the mechanical resonator design with the required resonance frequency.
A.R. defined the required high reflective optical coatings for the retro-reflection mirror and optical resonator.
A.R. designed and implemented the fibre coupled optical resonator.
A.R. developed the optical readout scheme to perform acceleration and displacement measurements using the optomechanical resonator.
Sv.Ab. and A.H. contributed to the optical layout and the electronic control and data acquisition system.
K.S. and D.T. performed the atom interferometry measurements and provided normalised atom population data for hybridisation.
A.R. hybridised the atom interferometer and optomechanical resonator signals via optimised vibration noise post-correction and performed data evaluation to analyse their performance individually and in hybridised configuration.
A.R. drafted the initial manuscript and D.S. contributed to and edited the final version.
All authors critically reviewed and approved of the final version.

\end{document}